\documentstyle[twoside,fleqn,espcrc2]{article}
\newcommand{\ttbs}{\char'134}
\newcommand{\AmS}{{\protect\the\textfont2
  A\kern-.1667em\lower.5ex\hbox{M}\kern-.125emS}}
\begin{document}

\title{Quantum Theory with Many Degrees 
of Freedom from Monte Carlo Hamiltonian}

\author{Xiang-Qian Luo
\address{Department of Physics,
       Zhongshan University, Guangzhou 510275, 
       China \\
and
National Center for Theoretical Sciences, 
Hsinchu, Taiwan 30043, China},
C.Q. Huang
\address{Department of Physics, Foshan Science 
and Technology College, Foshan 528000, China},
J.Q. Jiang
\address{Department of Physics, Guangdong Institute of Education, 
Guangzhou 510303, China},
H. Jirari $^{\rm{c}}$, H. Kr\"oger
\address{D\'epartement de Physique, Universit\'e Laval, 
Qu\'ebec, Qu\'ebec G1K 7P4, Canada},
K. J.M. Moriarty
\address{Department of Mathematics, 
Statistics and Computational Science, 
Dalhousie University, Halifax, Nova Scotia B3H 3J5, Canada}
}

\begin{abstract}
With our recently proposed effective Hamiltonian via Monte Carlo, 
we are able to compute low energy physics of
quantum systems. The advantage is that we can obtain not only 
the spectrum of ground and excited states, 
but also wave functions. 
The previous work has shown the success of this method 
in (1+1)-dimensional quantum mechanical systems. 
In this work we apply it to higher dimensional systems.
\end{abstract}

\maketitle


\section{Introduction}
Monte Carlo (MC) method with importance sampling is an excellent technique 
to calculate high dimensional integrals, and has successfully been applied 
to Lagrangian lattice gauge theory. 
Recently, we proposed an algorithm to compute the
MC Hamiltonian \cite{mch}.
The advantage, in comparison with the standard Lagrangian MC approach, 
is that one can obtain not only the spectrum of ground 
and excited states, but also the wave functions. 
The method has been tested in quantum mechanics (QM) in 1+1 dimensions, where
the exact results were reproduced with high precision 
\cite{mch,lat99_2,Huang}.
In this paper, we apply this algorithm to QM in 2+1 dimensions. 
The motivation is as follows.
(a) A number of algorithms for solving the Schr\"odinger equation 
in lower dimensions (e.g. Runge-Kutta) 
do not work in 2+1 dimensions. 
Our method works in 2+1 dimensions,
it is likely to work also in higher dimensions.
(b) The ultimate goal of the algorithm is 
to study structure function, the $S$ matrix, QCD at finite density 
as well as systems with many degrees of freedom. 
QM in 2+1 dimensions is a simple example.

\section{Algorithm}
Let's review briefly the basic ideas \cite{mch}. 
According to Feynman's path integral approach to QM, 
the (imaginary time) transition amplitude between an initial state at position  
$x_i$, and time $t_i$,  
and final state at $x_f$, $t_f$ is related 
to the Hamiltonian $H$ by
\begin{eqnarray*}
<x_{f},t_{f} | x_{i},t_{i}> 
= <x_{f} | e^{-H(t_{f}-t_{i})/\hbar} | x_{i}>
\end{eqnarray*}
\begin{eqnarray}
= \sum_{\nu=1}^{\infty} <x_{f} | E_{\nu} > 
e^{-E_{\nu} T/\hbar} < E_{\nu} | x_{i}>,
\end{eqnarray}
where $T=t_f-t_i$.
The starting point of our method, as described in detail in \cite{mch}
is to construct an effective Hamiltonian $H_{eff}$ 
(finite $N \times N$ matrix) by 
\begin{eqnarray*}
<x_{f} | e^{-H(t_{f}-t_{i})/\hbar} | x_{i}> \approx 
<x_{f} | e^{-H_{eff} T/\hbar} | x_{i}>
\end{eqnarray*}
\begin{eqnarray}
=\sum_{\nu=1}^{N} < x_{f} | E^{eff}_{\nu}> e^{-E^{eff}_{\nu} T/\hbar} 
< E^{eff}_{\nu} | x_{i} > .
\end{eqnarray}
$H_{eff}$ can be found by MC simulation 
using the following procedure: 

\noindent
(a) Discretize the continuous time.

\noindent
(b) Generate configurations $[x]$ obeying the Boltzmann distribution
\begin{eqnarray}
P(x) = { \exp(- S[x]/ \hbar)  \over
       \int [dx] \exp(- S[x]/ \hbar)}.
\end{eqnarray}
\noindent
(c) Calculate the transition matrix elements 
\begin{eqnarray}
M_{fi} = 
<x_{f} | e^{-H_{eff} T/\hbar} | x_{i}>
\end{eqnarray}
between $N$ discrete $x_i$ points and $N$ $x_f$ points. 
Note that the matrix $M$ is symmetric.

\noindent
(d) Diagonalize $M$ by a unitary transformation
\begin{eqnarray}
M &=& U^{\dagger}DU,
\end{eqnarray}
where 
$D =diag (e^{-E^{eff}_{1}T/\hbar},..., e^{-E^{eff}_{N}T/\hbar})$. 
Steps (a) and (b) are the same as the standard MC method.
Step (c) is the essential ingredient of our method, from which
we can construct $H_{eff}$, and obtain
the eigenvalues $E^{eff}_{\nu}$ and wavefunction 
$\vert E^{eff}_{\nu} >$ 
through step (d). Once the spectrum
and wave functions are available, all physical information can be retrieved.
Since the theory described by $H$ is now approximated by a theory
described by a finite matrix $H_{eff}$, the physics
of $H$ and $H_{eff}$ might be quite different at high energy. 
Therefore we expect that we can only
reproduce the low energy physics of the system.
This is good enough for our purpose.

\section{Results}
We consider the following examples in $D=2$:

\subsection{Uncoupled harmonic oscillator}
The Euclidean action is given by
\begin{eqnarray}
S={\int_{0}^{T}{dt}\left[{m \over 2}(\dot{x}^2+\dot{y}^2)
+{m{\omega}^2 \over 2} (x^2+y^2) \right]}.
\end{eqnarray}
The spectrum is degenerate and exactly known:
\begin{eqnarray}
E_{n_1,n_2}=\omega \hbar (n_1+n_2 +1),~n_1,n_2 = 0,1,2,\ldots
\label{Uncoupled}
\end{eqnarray}

\subsection {Coupled harmonic oscillator}   
The Euclidean action is given by 
\begin{eqnarray*}
S={\int_{0}^{T}{dt}\left[{m \over 2}(\dot{x}^2+\dot{y}^2)+{m{\omega}^2 
\over 2}(x^2+y^2)
+ \lambda x y \right]},
\end{eqnarray*}
\begin{eqnarray}
\rm{for} ~ 0<\lambda < m \omega^2.
\end{eqnarray}
The spectrum is also exactly known:  
\begin{eqnarray*}
E_{n_1,n_2}=\omega_1 \hbar (n_1+{1\over 2})
+\omega_2 \hbar (n_2+{1\over 2}),
\end{eqnarray*}
\begin{eqnarray}
n_1,n_2= 0,1,2,\ldots,
\end{eqnarray}
where $\omega_1=\sqrt{\omega^2+{\lambda \over m}}$ and 
$\omega_2=\sqrt{\omega^2-{\lambda \over m}}$.

\subsection{Summary}
Tabs. 1 and 2 compare the spectrum from $H_{eff}$
with the exact results.
Figs. 1-4 show the first two wave functions.
They are in very good agreement with the exact ones, 
at least in the low energy domain. 
In Ref. \cite{Jiang}, more results for wave functions 
of higher execited states,
and thermodynamical quantities 
such as the average energy and specific heat, were
reported.
However, the CPU time is much longer 
than in 1-dimensional systems. 
We need to find a better (stochastic) way to
select the basis in Hilbert space in step (c) of Sect. 2,
before extending the method to quantum field theory.
We are curious to see if 
our method can resolve the long standing problem 
of QCD at finite density.
It is worth mentioning the recent analytical \cite{Luo} 
and numerical  \cite{creutz} efforts in Hamiltonian formulation.

X.Q.L. is supported by the
National Science Fund for Distinguished Young Scholars,
supplemented by 
National Natural Science Foundation, 
fund for international cooperation and exchange,
Ministry of Education, 
and Hong Kong Foundation of
Zhongshan University Advanced Research Center. 
We also
thank the Lat'99 organizers and NCTS, Taiwan
for partial support.
H.K. would like to acknowledge support by NSERC Canada.

\begin{table}[hbt]
\caption{Spectrum of {\it uncoupled} hamonic oscilator.}
\begin{center}

\vspace{-15mm}
\caption{$\psi_{n_1n_2}(x,y)$ vs. $x$ at $y=-1$ 
for {\it uncoupled} harmonic oscillator with $n_1=n_2=0$.}
\label{fig1}
\end{figure}

\begin{figure}[htb]
\hspace{6mm}
\setlength{\unitlength}{0.240900pt}
\ifx\plotpoint\undefined\newsavebox{\plotpoint}\fi
\sbox{\plotpoint}{\rule[-0.200pt]{0.400pt}{0.400pt}}%
%
\vspace{-15mm}
\caption{$\psi_{n_1n_2}(x,y)$ vs. $x$ at $y=-1$
for {\it uncoupled} harmonic oscillator with $n_1=0, n_2=1$.}
\label{fig2}
\end{figure}

\begin{table}[hbt]
\caption{Spectrum of {\it coupled} hamonic oscilator.}
\begin{center}
%
\vspace{-15mm}
\caption{
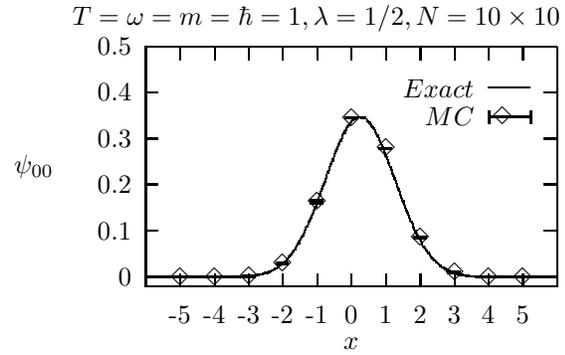
$\psi_{n_1n_2}(x,y)$ vs. $x$ at $y=-1$
for {\it coupled} harmonic oscillator with $n_1=n_2=0$.}
\label{fig3}
\end{figure}

\begin{figure}[htb]
\hspace{6mm}
\setlength{\unitlength}{0.240900pt}
\ifx\plotpoint\undefined\newsavebox{\plotpoint}\fi
\sbox{\plotpoint}{\rule[-0.200pt]{0.400pt}{0.400pt}}%
%
\vspace{-15mm}
\caption{$\psi_{n_1n_2}(x,y)$ vs. $x$ at $y=-1$
for {\it coupled} harmonic oscillator with $n_1=0, n_2=1$.}
\label{fig4}
\end{figure}

\end{document}